\documentclass[onecolumn]{IEEEtran}

\usepackage[T1]{fontenc}
\usepackage[utf8]{inputenc}
\usepackage{amsmath,amssymb}
\usepackage{booktabs}
\usepackage{array}
\usepackage{url}
\usepackage{tikz}
\usepackage{pgfplots}
\pgfplotsset{compat=1.17}
\usetikzlibrary{arrows.meta,positioning,fit,calc}

\usepackage{cleveref}

\newcommand{\pct}[1]{#1\,\%}

\begin{document}

\title{Cross-Platform Application Migration to HarmonyOS:
An Agentic Pipeline with Artifact-Derived Verification}

\author{%
\IEEEauthorblockN{Ergun Bi\c{c}ici$^{\dagger}$, Mehmet Kaygusuz,
Fatih Turan G\"undo\u{g}du, Emre Filiz, \c{C}a\u{g}atay K{\i}z{\i}lda\u{g},
B\"unyamin Eymen Alag\"oz, Emincan \"Ozcan, Sadi \c{C}elik,
Tolunay \"Ozt\"urk, Ege Bulut, Said Aldemir}\\[3pt]
\IEEEauthorblockA{Intelligent Application Development\\
Huawei Turkey R\&D Center, Istanbul, Turkey\\
\url{ergun.bicici@huawei.com}}
\thanks{$^{\dagger}$Corresponding author. 
}}

\maketitle

\begin{abstract}
The migration of applications written for a different platforms like Android or iOS to HarmonyOS NEXT requires a complete rewrite in ArkTS. Although LLMs automate the translation of source constructs into ArkTS, syntactic compilation is not enough and functional assurance is a critical open challenge in HarmonyOS NEXT migration. We present HMOS~Migrator, an agentic pipeline that reads an application's source code, generates an ArkTS project, and validates the result by replaying recorded interactions on a real device. We cast the scoring problem as quality estimation without a reference implementation; we score a migration using only the source application's recorded behaviour and the artifacts produced by the pipeline. We define seven weighted metrics, each constructed so that reduced measurement cannot artificially raise the score. Over twelve runs across six open-source applications, the pipeline's structural reproduction is high (completeness \pct{83.3}--\pct{100.0}), while on-device behaviour (\pct{4.3}--\pct{63.1}) and UI fidelity (\pct{21.1}--\pct{63.2}) remain the primary limits. Scores improve steadily over time: $+0.45$~points per day overall ($R^2=0.44$, $p=0.019$) and $+0.47$ over the model-free subset ($R^2=0.39$, $p=0.029$), with corresponding to fit angles of $+8.4^{\circ}$ and $+11.7^{\circ}$ respectively.
\end{abstract}


\section{Introduction}

HarmonyOS NEXT applications use the ArkTS and ArkUI programming model~\cite{arkts} rather than the Android runtime. Consequently, an existing Android or iOS application must be reconstructed for the target ecosystem rather than simply repackaged. Manual migration requires substantial effort across application logic, data access, UI construction, resources, navigation, and platform services. Large language models make this reconstruction amenable to automation, but application-scale migration also requires systematic execution and validation.

We present HMOS~Migrator, a fully automated agentic pipeline that carries an application from source repository to an installable HarmonyOS application and an artifact-derived verification report. In the measured NetNewsWire run, $117.8\times10^{3}$ source lines were processed in 17.9~h wall-clock time, equivalent to approximately 15.2~h per $100\times10^{3}$ source lines. Thus, an application of this scale can be migrated within a working day rather than through a multi-day manual port. The pipeline also records model usage and execution evidence, allowing both migration cost and output quality to be measured reproducibly.

Our second contribution is an artifact-derived verification scheme for a setting in which no independently implemented target application is available as a reference. The source application supplies recorded states and interaction traces, while the migration run supplies target artifacts and execution evidence. Deterministic programs recompute published quantities directly from these artifacts, while model-based methods are used only for semantic or visual judgements.
The resulting score is reported together with its evidence coverage rather than treating unobserved states as successful observations.
We cast scoring as \emph{quality estimation without a reference implementation}.
No correct port exists to compare against. We have the source application, the
states recorded from it, and the evidence the run leaves behind. 

We evaluate twelve migrations of six open-source applications and report both end-to-end verification results and measured execution cost. The evaluation demonstrates autonomous, application-scale migration from both Android and iOS, quantifies the associated time and token costs, and assesses the reproducibility and coverage of artifact-based verification.
\label{sec:intro}

This work makes three primary contributions: (1) the migration pipeline (Section~\ref{sec:pipeline}); (2) a verification framework demonstrating how naive metrics can reward reduced measurement (Section~\ref{sec:verify}); and (3) an end-to-end evaluation spanning twelve runs across six applications, uniformly rescored under a single code state (Sections~\ref{sec:experiments}).

\section{Related Work}
\label{sec:related}

The closest prior work evaluates ArkTS generation at the prompt level rather than application scale. Erku\c{s} et al.~\cite{arktseval} introduce a 300-prompt ArkTS benchmark covering 21 models and report low functional correctness, with the best model achieving \pct{22.7} Pass@1; performance is lower on UI tasks than on algorithmic tasks. Aytekin et al.~\cite{arkbaselines} propose Iterative Compilation Feedback, where models iteratively repair generated code using compiler feedback, increasing \emph{syntactic} Pass@1 to \pct{91}. 

UI migration has also been studied through rule-based transpilation~\cite{guimigrator}, which translates code horizontally between readable languages. These approaches target platforms with established correspondences between source and target UI frameworks. Ours has none, so each screen must be reconstructed rather than rewritten. Nor is there a correct target implementation to compare the result against, which is why quality must be estimated from the source application's own recorded behaviour. Additionally, visual layout matching alone cannot verify functional correctness, as a visually accurate screen may still fail to operate. Nor is a language model a dependable translator: Pan et al.~\cite{lostintranslation} catalogue the bugs introduced when models translate code between languages, which is the class of defect our verification exists to expose. Interaction-trace transfer provides a more relevant benchmark for our validation stage~\cite{craftdroid,mapit}, yet prior work assumes both applications are already fully implemented. In our setting, the target app is dynamically synthesized during migration. Replay thus fulfills two essential roles: exercising the generated codebase and producing trace data for cross-application behavioral comparison.

Quality estimation without a reference is well established in machine translation. Referential Translation Machines~\cite{rtm2013} estimate translation quality without access to a reference translation, while Bi\c{c}ici~\cite{mtpps2022} predicts translation performance from the source alone. Our setting differs from both: there is no independently implemented target application, but recorded source states are available and replayed target states can be compared with them directly. We therefore adopt the quality-estimation perspective rather than its predictive machinery. In particular, our estimator does not train a quality predictor or extract predictive features; instead, it derives scores from execution and UI artifacts and requires the score to remain sensitive to measurement coverage. The temporal trend analysis uses the normalized fit and angle statistic of~\cite{bicici2021}.

\section{Pipeline Architecture}
\label{sec:pipeline}

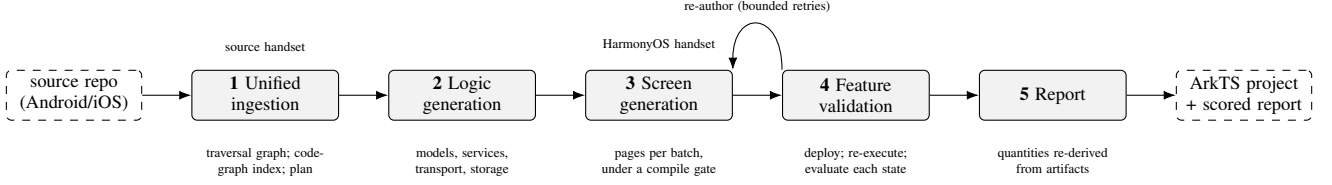
\begin{figure}[t]
\centering
\begin{tikzpicture}[
  font=\scriptsize,
  box/.style={draw,rounded corners=2pt,align=center,minimum height=6.6mm,
              inner sep=2pt,text width=18mm},
  io/.style={draw,dashed,rounded corners=2pt,align=center,minimum height=6.6mm,
             inner sep=2pt,text width=16.5mm},
  ph/.style={box,fill=black!5},
  a/.style={-{Latex[length=1.6mm]}},
]
\node[io] (src) {source repo\\(Android/iOS)};
\node[ph,right=6.5mm of src] (p1) {\textbf{1} Unified\\ingestion};
\node[ph,right=6.5mm of p1] (p2) {\textbf{2} Logic\\generation};
\node[ph,right=6.5mm of p2] (p3) {\textbf{3} Screen\\generation};
\node[ph,right=6.5mm of p3] (p4) {\textbf{4} Feature\\validation};
\node[ph,right=6.5mm of p4] (p5) {\textbf{5} Report};
\node[io,right=6.5mm of p5] (out) {ArkTS project\\+ scored report};
\foreach \a/\b in {src/p1,p1/p2,p2/p3,p3/p4,p4/p5,p5/out}
  \draw[a] (\a) -- (\b);
\node[below=2.4mm of p1,text width=24mm,align=center,font=\tiny]
  {traversal graph; code-graph index; plan};
\node[below=2.4mm of p2,text width=24mm,align=center,font=\tiny]
  {models, services, transport, storage};
\node[below=2.4mm of p3,text width=24mm,align=center,font=\tiny]
  {pages per batch, under a compile gate};
\node[below=2.4mm of p4,text width=24mm,align=center,font=\tiny]
  {deploy; re-execute; evaluate each state};
\node[below=2.4mm of p5,text width=24mm,align=center,font=\tiny]
  {quantities re-derived from artifacts};
\node[above=1.2mm of p1,font=\tiny] {source handset};
\node[above=1.2mm of p3,font=\tiny] {HarmonyOS handset};
\draw[a] (p4.north west) .. controls +(0,8mm) and +(0,8mm) .. (p3.north east)
  node[midway,above=0.3mm,font=\tiny] {re-author (bounded retries)};
\end{tikzpicture}
\caption{The five phases, the handset each requires, and the bounded repair loop.
A state still divergent at the retry bound is labelled as requiring review and
execution proceeds.}
\label{fig:pipeline}
\end{figure}

The system is by one command: /migrate. The tool accepts a source repository and two connected devices—one executing the original application and the other running HarmonyOS—and produces an ArkTS project alongside a migration report. It can also use a corresponding emulator if it exists such as the Android Studio for Android and DevEco Studio for HarmonyOS.

The system consists of five phases (Fig.~\ref{fig:pipeline}):
\begin{enumerate}
    \item \emph{Phase 1} \emph{unified ingestion}, which extracts source structure and runtime states; uses stable element identifiers, drives the source application to reach every navigable screen, and records a traversal graph whose nodes carry a screenshot and an index of the code repository. A node is a \emph{state}: a screen plus the taps that
reached it. The same screen reached two ways gives two states, so states
outnumber screens. 
\item \emph{Phase 2} \emph{logic generation}, which generates the code for the models, services, storage, and application logic; 
\item \emph{Phase 3} \emph{screen generation}, which produces ArkUI pages under compilation gates; 
\item \emph{Phase 4} \emph{feature validation}, which deploys the application and replays recorded interactions on HarmonyOS hardware or emulator; evaluates each source/target state pair across multiple layers and returns divergent pairs for re-authoring within a bounded retry budget.
\item \emph{Phase 5} \emph{reporting}, which derives verification quantities from the artifacts produced by the run.

\end{enumerate}

\section{Verification Scheme}
\label{sec:verify}

\begin{table}[t]
\caption{The seven metrics for verification.}
\label{tab:metrics}
\centering
\footnotesize
\begin{tabular}{@{}rlcl@{}}
\toprule
\# & Metric & Weight & LLM judge\\
\midrule
1 & Screen capture coverage          & 6  & no\\
2 & Migration completeness (breadth) & 17 & no\\
3 & Navigation edges                 & 12 & no\\
4 & Feature verify (on-device)       & 27 & \textbf{yes} --- one verdict per check\\
5 & Replay coverage                  & 11 & no\\
6 & Extended-tree vision comparison  & 7  & \textbf{yes} --- one verdict per state\\
7 & UI fidelity (element overlap)    & 20 & no\\
\midrule
\multicolumn{2}{@{}l}{\textsc{overall} (all steps)}    & 100 & 34 of 100 weight judged\\
\multicolumn{2}{@{}l}{\textsc{overall} (no LLM judge)} & 66  & steps 1, 2, 3, 5, 7\\
\bottomrule
\end{tabular}
\end{table}

The verification report contains 7 metrics (Table~\ref{tab:metrics}), each based on artifacts that the output stores and their scores are weighted. The overall score becomes:
$$ S=\sum_{i=1}^{7} w_i s_i,\qquad \sum_i w_i=1 $$

where every \(s_i\in[0,1]\) with weights

$$ \mathbf w=(0.06,0.17,0.12,0.27,0.11,0.07,0.20). $$

\smallskip\noindent\textbf{1. Screen capture coverage} (6\%, deterministic).\enspace The share of the source screen inventory that source analysis actually reached and photographed. 
Let $\mathcal{S}$ be the source screen inventory and $\mathcal{R}\subseteq\mathcal{S}$ the screens reached while exploring the source application, where a screen is considered reached when at least one of its states produces a valid capture:
\begin{equation}
s_1=\frac{|\mathcal{R}|}{|\mathcal{S}|}
\end{equation}
\emph{An uncaptured screen cannot be recovered downstream:} without a reference image, it cannot be planned, replayed, or evaluated, so $s_1$ limits all subsequent metrics.

\smallskip\noindent\textbf{2. Migration completeness} (17\%, deterministic).\enspace The ratio of distinct source screens converted to the total set of screens in the original application.
Let $\mathcal{S}$ be the set of distinct source screens, and let $\mathcal{P}\subseteq\mathcal{S}$ denote those realized by generated pages with actual content. A page that compiles and supports navigation but renders no content is classified as an unpopulated layout placeholder and excluded from $\mathcal{P}$.
\begin{equation}
s_2=\frac{|\mathcal{P}|}{|\mathcal{S}|}
\end{equation}
This measures breadth of migration, not whether the implementation actually works.

\smallskip\noindent\textbf{3. Navigation edges} (12\%, deterministic).\enspace The share of recorded source transitions reproduced in the target. \emph{Blind to whether either endpoint works:} it establishes only that the screens are wired to one another. 
Let $E_m$ be the mappable recorded source transitions whose two endpoints have generated pages and are distinct. Let $E_r\subseteq E_m$ be the reproduced transitions that the generated code actually performs, either through a direct route or a data-driven route that resolves to the destination.

\begin{equation}
s_3=\frac{|E_{r}|}{|E_{m}|}
\end{equation}
A route table existing is not a wired edge: $E_{r}$ counts the pushes the pages actually perform. This is deliberately different from screen coverage: having a screen does not imply that the user can reach it correctly.

\smallskip\noindent\textbf{4. Feature verify} (27\%, model-judged).\enspace For each captured state, every feature attributed to that screen is evaluated using its evidence bundle: device logs, UI trees, before-and-after gesture dumps. So one check is one feature's verdict on one state. Checks supported by sufficient evidence form the exercised set $E$, with $V\subseteq E$ denoting those that are verified. A check is critical when the feature it names is rated high-risk in the source inventory, or when a majority of the screens using that feature are; promoting on any single high-risk screen instead leaves almost every feature critical. The denominator also includes $U_c$, the unexercised checks that are either critical or have no catalogued feature, and $O$, the high-risk features not covered by any check (orphans):
\begin{equation}
s_4=\frac{|V|}{|E|+|U_c|+|O|}
\end{equation}
$U_c$ and $O$ enter the denominator only, so each can only lower the score. That is what stops the metric from being improved by deleting links: a check that loses its link moves into $U_c$, and a feature that loses its last check moves into $O$, so the denominator never shrinks. Features that cannot be verified in this environment are excluded. When fewer than ten checks are exercised, the score falls back to $|V|$ over all checks. Feature verify is the only metric that shows a control actually runs and does what it should, rather than merely appearing in the interface.

\smallskip\noindent\textbf{5. Replay coverage} (11\%, deterministic).\enspace The share of source states confirmed reached on the corresponding target screen when the recorded interaction paths are re-executed. Measures how much of the recorded source interaction trace can actually be executed on the target. Re-execution filters the captured source states $S$ into the \emph{attempted} $A$ (replay drove its recorded path), the \emph{captured} $C_{\mathrm{cap}}$ (the attempt returned a screenshot) and the \emph{confirmed} $C$ (of the intended screen).
\begin{equation}
s_5=\frac{|C|}{|S|},\qquad C\subseteq C_{\mathrm{cap}}\subseteq A\subseteq S
\end{equation}
The denominator remains $|S|$ rather than $|A|$ so that states never replayed are counted as uncovered rather than excluded from the metric.
This should remain distinct from feature correctness. A button may be reachable and replayable but still perform the wrong operation. It bounds metrics 4, 6 and 7 as metric 1 bounds everything.

\smallskip\noindent\textbf{6. Extended-tree vision} (7\%, model-judged).\enspace A model compares each captured target state with its source reference and gives a graded verdict. Three values score: \emph{ok}, no defect; \emph{minor}, a medium or low defect or an item the judge could not see; and \emph{fail}. States that are not captured or judged are excluded from scoring, but still counted.

$$ s_6=\frac{n_{\mathrm{ok}}+0.90\,n_{\mathrm{minor}}}{n_{\mathrm{ok}}+n_{\mathrm{minor}}+n_{\mathrm{fail}}}$$

\smallskip\noindent\textbf{7. UI fidelity} (20\%, deterministic).\enspace The pooled $F_1$ score evaluates elements present across paired source and target accessibility tree dumps. Strictly deterministic, this metric measures matching semantic labels rather than raw pixel outputs. For $M$ pooled matched elements, $S$ labeled source elements, and $T$ labeled target elements, the score is defined as:

\[
P=\frac{M}{T},\qquad R=\frac{M}{S},\qquad
s_7=\frac{2PR}{P+R}=\frac{2M}{S+T}
\]

\section{Experiments}
\label{sec:experiments}

\begin{table}[t]
\caption{Subject applications and the versions migrated. \emph{Screens} is the source inventory; a range spans that application's runs.}
\label{tab:apps}
\centering
\footnotesize
\begin{tabular}{@{}lllrr@{}}
\toprule
Application & Src & Version & Screens & Runs\\
\midrule
VLC             & Android & 3.7.1               & 49--149 & 5\\
NewPipe         & Android & 0.28.8              & 66 & 1\\
Wikipedia       & Android & 50598-r, 50599-beta & 193, 223 & 2\\
NetNewsWire     & iOS     & 7.1.3               & 28 & 2\\
Image Toolbox   & Android & 4.1.0               & 79 & 1\\
SwiftRadio      & iOS     & 3.0.0               & 9 & 1\\
\bottomrule
\end{tabular}
\end{table}

Twelve runs across six open-source applications are reported in Table~\ref{tab:apps}. 
For each run, the version is taken from its own vendored checkout, i.e., the exact code ingested by the pipeline. 
We score, publish and fit with a performance trend fit line~\cite{bicici2021} all twelve runs.

\subsection{Model generation is not constant}
\label{sec:models}
The release of Claude Opus~5 on 2026-07-24 occurred midway through data collection, introducing model variation into the corpus. Model assignment is strictly based on active execution windows: three pre-release runs (VLC 14/07, NewPipe 20/07, Wikipedia 22/07) are attributed to Opus~4.8, while seven subsequent runs belong to Opus~5. A single run (VLC 29/07) spans the release boundary and is annotated separately. The final run (SwiftRadio 28/08) was executed with GPT~5.6 Luna, so the corpus is not a single-vendor series either. This also influences performance trends across time.

\subsection{Execution cost}

Table~\ref{tab:cost} summarizes the empirical execution costs for the NetNewsWire run (17/08), covering $117.8\times10^{3}$ source lines across 882 files. We distinguish between three distinct temporal metrics: \emph{wall-clock time} (17.9~h, including idle durations), \emph{active wall-clock time} (13.8~h, omitting idle periods and flattening concurrent subagent execution into single intervals), and cumulative \emph{agent work-hours} (16.9~h, aggregating each subagent's runtime individually and thus exceeding active wall-clock due to parallel authoring).
Token consumption is overwhelmingly dominated by prompt caching. Of the $1.87\times10^{9}$ total processed tokens, $1.82\times10^{9}$ (\pct{97.3}) represent repeated context re-reads resubmitted across turns, whereas the model generated only $5.23\times10^{6}$ output tokens. 

\begin{table}[t]
\caption{Instrumented cost of the NetNewsWire 17/08 run (2026-08-17 17:31--2026-08-18 11:27 UTC). Work-hours measure agent effort. \emph{Processed} tokens comprise input, cache reads, and output; \emph{new input} excludes cache reads. Percentages give each phase's share of the run total for the corresponding column, except \emph{cache-read}, whose percentage is the phase's cache-read share of its own processed tokens. Percentages use the denominator specified for each column, so the phase shares differ across columns.}
\label{tab:cost}
\centering
\footnotesize
\begin{tabular}{@{}lrrrrrrrrr@{}}
\toprule
Phase & \multicolumn{2}{c}{Work-hours} & \multicolumn{2}{c}{Processed}
      & \multicolumn{2}{c}{Cache-read} & \multicolumn{2}{c}{New input} & Gen.\\
\cmidrule(lr){2-3}\cmidrule(lr){4-5}\cmidrule(lr){6-7}\cmidrule(lr){8-9}
      & h & \% & tokens & \% & tokens & \% & tokens & \% & tokens\\
\midrule
1 --- ingestion          & 3.43 & 20.3 & 295.5\,M & 15.8 & 282.9\,M & 95.7 & 11.55\,M & 25.4 & 998\,K\\
2 --- logic generation   & 0.95 &  5.6 & 182.8\,M &  9.8 & 181.4\,M & 99.3 &  1.07\,M &  2.4 & 252\,K\\
3 --- screen generation  & 6.34 & 37.4 & 623.1\,M & 33.3 & 609.0\,M & 97.7 & 12.21\,M & 26.8 & 1.88\,M\\
4 --- feature validation & 6.22 & 36.7 & 771.1\,M & 41.2 & 748.3\,M & 97.0 & 20.67\,M & 45.4 & 2.11\,M\\
\midrule
Total                    & 16.9 & 100  & 1.87\,B  & 100  & 1.82\,B  & \textbf{97.3} & 45.50\,M & 100 & 5.23\,M\\
\bottomrule
\end{tabular}
\\[3pt]
{\footnotesize Wall-clock 17.9\,h $\cdot$ active wall-clock 13.8\,h $\cdot$ 86
subagents $\cdot$ 1{,}145 main-thread turns.}
\end{table}

\subsection{Results}
\label{sec:results}

\begin{table}[t]
\caption{Results over seven metrics. All VLC rows
are television-excluded. \emph{Screens} is the source screen inventory prior to that exclusion; \emph{Model} is attributed per Section~\ref{sec:models}.}
\label{tab:results}
\centering
\footnotesize
\setlength{\tabcolsep}{4pt}
\begin{tabular}{@{}llclr rrrrrrr rr@{}}
\toprule
& & & & & \multicolumn{7}{c}{Metric score (\%)} & \multicolumn{2}{c}{Aggregate}\\
\cmidrule(lr){6-12}\cmidrule(l){13-14}
Run & Application & Src & Model & Screens
& 1 & 2 & 3 & 4 & 5 & 6 & 7 & all & no-LLM\\
\midrule
VLC 14/07/2026              & VLC          & Android & Opus 4.8  & 49  & 29.2 & 83.3 & 50.0 & 28.3 & 2.9   & 94.4 & 48.0 & 46.1 & 48.2\\
NewPipe 20/07/2026          & NewPipe      & Android & Opus 4.8  & 66  & 75.8 & 100.0& 87.5 & 33.3 & 33.6  & 95.7 & 61.8 & 63.8 & 72.9\\
Wikipedia 22/07/2026        & Wikipedia    & Android & Opus 4.8  & 193 & 34.2 & 99.5 & 78.6 & 21.1 & 75.0  & 93.1 & 49.8 & 58.8 & 70.6\\
Wikipedia 29/07/2026        & Wikipedia    & Android & Opus 5    & 223 & 31.8 & 99.5 & 100.0& 4.3  & 92.9  & 71.4 & 31.8 & 53.6 & 71.8\\
VLC 29/07/2026              & VLC          & Android & 4.8$\to$5 & 49  & 100.0& 87.5 & 73.5 & 28.3 & 12.6  & 87.0 & 21.1 & 49.0 & 53.5\\
VLC 02/08/2026              & VLC          & Android & Opus 5    & 149 & 91.2 & 100.0& 100.0& 12.0 & 88.7  & 36.3 & 22.6 & 54.5 & 73.9\\
VLC 06/08/2026              & VLC          & Android & Opus 5    & 96  & 79.2 & 90.6 & 93.2 & 15.2 & 75.3  & 66.2 & 56.3 & 59.6 & 77.1\\
NetNewsWire 16/08/2026      & NetNewsWire  & iOS     & Opus 5    & 28  & 96.4 & 96.4 & 100.0& 29.0 & 77.8  & 52.1 & 46.9 & 63.6 & 79.0\\
NetNewsWire 17/08/2026      & NetNewsWire  & iOS     & Opus 5    & 28  & 96.4 & 96.4 & 97.0 & 54.4 & 100.0 & 91.1 & 59.6 & 77.8 & 86.0\\
VLC 19/08/2026   & VLC          & Android & Opus 5    & 138 & 77.8 & 96.9 & 87.0 & 50.7 & 70.9  & 92.6 & 28.6 & 65.3 & 68.3\\
Image Toolbox 29/08/2026& Image Toolbox& Android & Opus 5    & 79  & 98.7 & 100.0& 100.0& 63.1 & 98.7  & 90.4 & 63.2 & \textbf{81.8} & \textbf{88.5}\\
SwiftRadio 28/08/2026   & SwiftRadio   & iOS     & GPT 5.6   & 9   & 100.0& 100.0& 77.8 & 30.4 & 100.0 & 85.5 & 50.0 & 67.5 & 80.8\\
\midrule
\multicolumn{5}{@{}l}{Weight (\%)} & 6 & 17 & 12 & 27 & 11 & 7 & 20 & 100 & 66\\
\bottomrule
\end{tabular}
\end{table}

Table~\ref{tab:results} summarizes the twelve runs across the seven verification metrics. 

Replay coverage is an important discriminator between weak and strong runs. Runs with low replay coverage provide limited evidence for downstream feature and visual verification, whereas higher coverage enables more source states to be exercised and evaluated. Thus, high structural completeness alone does not guarantee a high overall score: screens and navigation may be successfully reproduced, but their behaviour and visual fidelity cannot be fully assessed unless the corresponding states are reachable during replay.

Overall, the results suggest improvement over time. The aggregate scores tend to increase in the later runs, and the model-free aggregate shows a positive temporal slope. The improvement is also visible within some applications, particularly across the VLC and NetNewsWire runs, but it is not monotonic across individual metrics. Moreover, the model generation changes during the series and the applications differ in difficulty. 
\emph{Scores show a positive temporal trend.}
Across the twelve runs, the all-metric aggregate increases by $+0.450$ points per day ($R^2=0.44$, $p=0.0191$). The model-free aggregate increases by $+0.474$ points per day ($R^2=0.39$, $p=0.0292$). The corresponding normalized fit angles are $+8.4^{\circ}$ and $+11.7^{\circ}$. Thus, later runs tend to score higher, but the trend should be interpreted cautiously (\Cref{fig:trend}).

Three factors limit this interpretation. First, the sample contains only twelve runs: refitting without each run in turn leaves nine of twelve refits below the $0.05$ level, and the three that cross it---dropping VLC 14/07, Image Toolbox or SwiftRadio---show how much the result rests on individual observations. Second, the runs differ in both application and model generation; the series includes the transition from Opus~4.8 to Opus~5 and a final run executed with GPT~5.6 Luna. Third, the latest observation is a different application from the earlier runs. Consequently, the regression cannot distinguish improvement in the migration pipeline from differences in application difficulty or model capability. We therefore report the positive trend as an observed association rather than as evidence of demonstrated pipeline improvement.

\emph{Trend.} Developments in LLM models, such as the transition from Opus 4.8 to Opus 5 and the use of GPT 5.6 Luna on the final run, together with iterative improvements to the HMOS Migrator tool, enable this observed improvement.

\providecommand{\measuredangle}{\angle}%
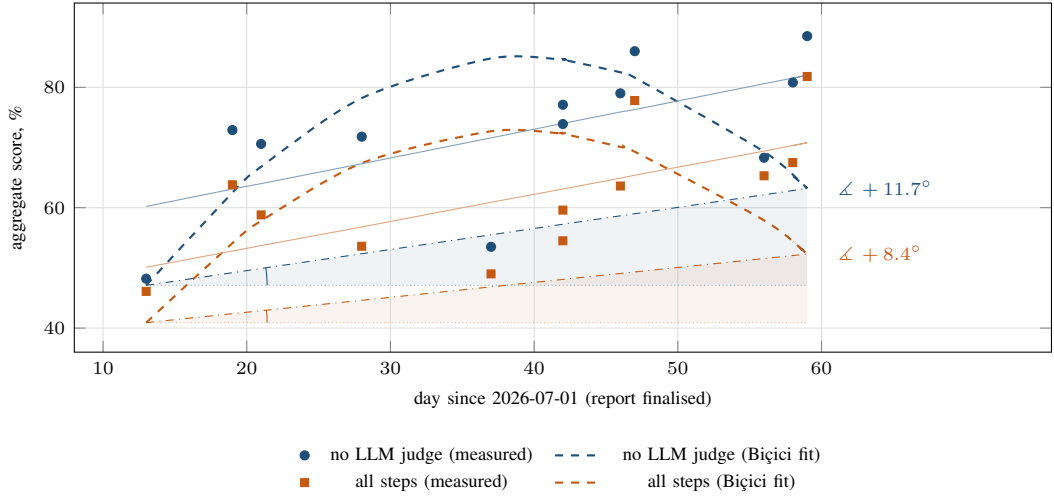
\begin{figure}[t]
\centering
\definecolor{trendA}{RGB}{31,78,120}
\definecolor{trendB}{RGB}{196,89,17}
\begin{tikzpicture}
\begin{axis}[
  width=0.80\textwidth, height=6.2cm,
  xlabel={day since 2026-07-01 (report finalised)},
  ylabel={aggregate score, \%},
  xmin=8, xmax=76, ymin=36, ymax=94,
  xtick={10,20,30,40,50,60},
  grid=major, grid style={black!12},
  tick label style={font=\scriptsize}, label style={font=\scriptsize},
  legend style={font=\scriptsize,at={(0.5,-0.24)},anchor=north,
                legend columns=2,draw=none,column sep=6pt},
]
\addplot[only marks,mark=*,mark size=1.7pt,trendA] coordinates {(13,48.2) (19,72.9) (21,70.6) (28,71.8) (37,53.5) (42,73.9) (42,77.1) (46,79) (47,86) (56,68.3) (59,88.5) (58,80.8)};
\addlegendentry{no LLM judge (measured)}
\addplot[smooth,dashed,thick,trendA] coordinates {(13,47.1) (19,62.5) (21,66.8) (28,78.2) (37,84.8) (42,84.6) (42,84.6) (46,82.5) (47,81.6) (56,69.3) (59,63.2) (58,65.3)};
\addlegendentry{no LLM judge (Bi\c{c}ici fit)}
\addplot[only marks,mark=square*,mark size=1.5pt,trendB] coordinates {(13,46.1) (19,63.8) (21,58.8) (28,53.6) (37,49) (42,54.5) (42,59.6) (46,63.6) (47,77.8) (56,65.3) (59,81.8) (58,67.5)};
\addlegendentry{all steps (measured)}
\addplot[smooth,dashed,thick,trendB] coordinates {(13,40.9) (19,54.1) (21,57.8) (28,67.4) (37,72.7) (42,72.3) (42,72.3) (46,70.1) (47,69.3) (56,57.9) (59,52.3) (58,54.3)};
\addlegendentry{all steps (Bi\c{c}ici fit)}
\addplot[trendA!55,forget plot] coordinates {(13,60.2) (19,63.1) (21,64) (28,67.3) (37,71.6) (42,74) (42,74) (46,75.9) (47,76.3) (56,80.6) (59,82) (58,81.5)};
\addplot[trendB!55,forget plot] coordinates {(13,50.1) (19,52.8) (21,53.7) (28,56.8) (37,60.9) (42,63.1) (42,63.1) (46,64.9) (47,65.4) (56,69.4) (59,70.8) (58,70.3)};
\path[fill=trendA,opacity=0.07] (axis cs:13,47.1) -- (axis cs:59,47.1)
  -- (axis cs:59,63.2) -- cycle;
\path[fill=trendB,opacity=0.07] (axis cs:13,40.9) -- (axis cs:59,40.9)
  -- (axis cs:59,52.3) -- cycle;
\addplot[densely dotted,trendA!45,forget plot] coordinates {(13,47.1) (59,47.1)};
\addplot[dash dot,trendA,forget plot] coordinates {(13,47.1) (59,63.2)};
\addplot[densely dotted,trendB!45,forget plot] coordinates {(13,40.9) (59,40.9)};
\addplot[dash dot,trendB,forget plot] coordinates {(13,40.9) (59,52.3)};
\draw[thin,trendA] let \p1=($(axis cs:59,63.2)-(axis cs:13,47.1)$),
        \n1={atan2(\y1,\x1)} in
  ($(axis cs:13,47.1)+(0:16mm)$) arc (0:\n1:16mm);
\draw[thin,trendB] let \p1=($(axis cs:59,52.3)-(axis cs:13,40.9)$),
        \n1={atan2(\y1,\x1)} in
  ($(axis cs:13,40.9)+(0:16mm)$) arc (0:\n1:16mm);
\node[font=\scriptsize,anchor=west,trendA] at (axis cs:60.5,63.2) {$\measuredangle\,+11.7^{\circ}$};
\node[font=\scriptsize,anchor=west,trendB] at (axis cs:60.5,52.3) {$\measuredangle\,+8.4^{\circ}$};
\end{axis}
\end{tikzpicture}
\caption{Both aggregates against report date, all twelve runs fitted. Dashed curves
are the Bi\c{c}ici fit~\cite{bicici2021}---$y/x$ regressed on $x$, unnormalised to
$f(x)=ax^{2}+bx$---and thin solid lines the least-squares slopes ($+0.474$ points per
day for the model-free lane, $p=0.0292$; $+0.450$ for all steps, $p=0.0191$). Each
dash-dotted chord joins its Bi\c{c}ici fit at the first and last observation: that
chord is the angle statistic, measured on axes normalised by 100 and by $x_{\max}$,
so the printed value is not the angle subtended here. The model-free lane sits above the
all-steps lane throughout, and both fits turn down to reach the rightmost
observations.}
\label{fig:trend}
\end{figure}\section{Conclusion}
\label{sec:conclusion}

An agentic pipeline carries an Android or iOS application to a compilable,
installable HarmonyOS application and replays its recorded interactions on real
hardware. The remaining difficulty is estimating how much of the result is correct
with no reference implementation to compare against. Across twelve runs and six
applications, structural migration is largely solved (completeness
\pct{83.3}--\pct{100.0}). On-device behaviour (\pct{4.3}--\pct{63.1}) and
element-level fidelity (\pct{21.1}--\pct{63.2}) are not. Score rises with date at
$+0.45$ to $+0.47$ points per day, statistically significantly with $p$ between $0.019$ and $0.029$. 

We presented an evidence-based framework for evaluating AI-driven application migration from Android and iOS platforms to HarmonyOS even though the underlying pipeline is platform independent. Across twelve runs, structural reproduction was generally strong, while functional verification and UI fidelity remained the main limitations. The results show that successful generation and compilation do not establish behavioural correctness: replay coverage and on-device evidence are essential for determining whether the generated application actually works. The framework also preserves unexercised and unverifiable cases rather than treating them as successes, making the resulting scores auditable and reproducible.

The aggregate results show a positive temporal trend, with a fitted increase of $+0.450$ points per day and a normalized angle of $+8.4^{\circ}$. The model-free aggregate increases by $+0.474$ points per day, corresponding to $+11.7^{\circ}$. These angles summarize the observed trend after normalizing the axes, but do not establish causal improvement because the corpus is small and spans different applications and model versions. Future work involves expanding the evaluation across more runs and time points and increase verification coverage to better separate improvements in generation capability from improvements in end-to-end migration quality.

\end{document}